\documentclass[12pt]{iopart}

\usepackage{iopams}
\usepackage{graphicx}
\usepackage[breaklinks=true,colorlinks=true,linkcolor=blue,urlcolor=blue,citecolor=blue]{hyperref}
\usepackage{amssymb,subfigure,color}
\usepackage{amsthm}

\renewcommand{\vec}[1]{\mbox{\boldmath$\mathrm{#1}$}}

\begin{document}

\title{Detection of magnetic impurities using electron vortex beams}

\author{Yan Wang}
\address{Institute of Modern Physics, Chinese Academy of Science,730000 Lanzhou, China}

\author{Chenglong Jia}
\address{Key Laboratory for Magnetism and Magnetic Materials of MOE, Lanzhou University, 730000 Lanzhou, China}

\author{Pengming Zhang}
\address{School of Physics and Astronomy, Sun Yat-sen University, 519082 Zhuhai, China}
\eads{\mailto{zhangpm5@mail.sysu.edu.cn}}

\begin{abstract}
Electron microscopy stands out as electron waves providing higher spatial resolving power compared to their optical counterpart. Here we investigate theoretically the interaction of twisted electrons generated in transmission electron microscope (TEM) and magnetic impurity, in which the magnetic dipole moment is taken as a demonstration element.  
In addition to the usual optical phase, the inhomogeneous vector potential generated by the magnetic dipole moment makes additional contribution to the intrinsic orbital angular momentum of electrons, resulting in a Gouy phase shift. By interfering the outgoing twisted electron beam with a reference Gaussian-cylindrical wave, one can determine the magnitude and orientation of magnetic dipole directly via the rotational and deformed interference pattern. The obtained results demonstrate the usefulness of twisted electron beams for probing the atomic- and nanoscale  magnetism of impurity by TEM and the proposed model provides the conceptual basis for future developments of the TEM method.
\end{abstract}



\maketitle

\section{Introduction}

Recently, twisted electron beams  attract more and more attentions so as to clarify the underlying physics and  the potential for novel technological applications \cite{Bliokh_2007,Uchida_2010,Verbeeck_2010,Mcmorran_2011,Schattschneider_2011,Bliokh_2011,Ivanov_2011,Karimi_2012,Lloyd_2012,Gallatin_2012,RHarvey2014,Ju2018,Pierce2019}. 
These electron vortex states carry a definite amount of intrinsic orbital angular momentum (OAM) $\ell$ and possess a helical phase front $e^{i\ell\phi}$. In experiment,  they can be produced in electron microscopes where the electrons are controlled and focused by using magnetic lenses. In contrast to the usual plane wave states, for which $\ell=0$, the OAM projection of twisted electrons can be as high as $\hbar\ell=1000\hbar$\cite{Mafakheri2017}. Such a huge magnetic moment resulting from large $\ell$ makes vortex beams particularly suitable for probing magnetic properties of materials at atomic/nano-scale \cite{Grillo2017a,Bliokh2017a,Lloyd2017a} and manipulating nanoparticles\cite{Gnanavel2012}. The OAM-induced moment also allows to enhance and explore magnetic phenomena in electron-light coupling\cite{Gallatin_2012,Bliokh2012,Ivanov2013}. Moreover, a nonzero OAM can significantly influence fundamental atomic and molecular collision processes\cite{Kosheleva2018,Maiorova2018}. The studies of the physical behavior of twisted electron beams in the presence of magnetic fields have been reported in recent years.\cite{Bliokh2012,Guzzinati2013} In the present work, an approach to detect magnetic impurity in materials by investigating the interaction of twisted electrons and the impurity-induced magnetic field is proposed.

The problem of magnetic impurities embedded in electron systems has attracted a lot of attention for many decades. Only taking Kondo effect as an example, magnetic impurities in metals induce manybody correlations which at low temperatures quench the spin fluctuations at the impurity site.\cite{Kondo effect} The magnetic structure of the impurities plays important roles of the Kondo screening. Therefore, it is important to explore the magnetic configuration of magnetic impurities. Magnetic dipoles are fundamental, observable units of magnetism and of great importance in the study of the magnetic properties of matter in static magnetic fields. Magnetic strength of the magnet source can be considered as a superposition of magnetic dipole moments with different magnitude and orientation. Therefore, the detail study of a magnetic dipole moment can provide fundamental and useful informations of magnetic  impurities. 

In contrast to the consideration that the magnetic dipole is comprised of two monopoles having opposite magnetic charges at a small distance from each other, and the twisted electrons interact with two magnetic monopoles separately\cite{Grillo2017a}, much less attention has been paid to the influence of the magnetic dipole as a whole on the propagating twisted electrons. Here, we propose the theoretical model describing the physical behavior of twisted electrons in the field generated by the magnetic dipole. Different from the uniform magnetic fields\cite{Bliokh2012}, the magnetic dipole field is inhomogeneous in space, so does the vector potential $\vec{A}$ ($\vec{B}=\nabla\times\vec{A}$). In our study, a special attention is paid to the phases of the outgoing electron wave. An Aharonov-Bohm phase\cite{Aharonov1959} is acquired as the transverse trajectory of twisted electron passing through the magnetic dipole and this results in an increase or decrease of the OAM, which is directly related to the Gouy phase shift\cite{Bliokh2012,Siegman1986}. Moreover, the magnetic field also gives rise to an electron optical phase shift. We show that the size of these shifts is dependent on the magnitude and orientation of magnetic dipole. The observation of these phase shifts by interfering the outgoing beam with a reference Gaussian-cylindrical wave highlights the potential of the proposed model for the detection of magnetic impurities using the twisted electron beams.

\section{Theoretical formulation}
As described theoretically by Nye and Berry \cite{Nye1974}, vortex waves are solutions of the three-dimensional (3D) wave equation with an angular-dependent phase factor of the form
\begin{equation}
\Psi(\vec{r})_{\ell}=\psi(\rho,z)e^{i\ell\phi},
\end{equation}
with $\rho$ and $\phi$ being the radial and azimuthal coordinates with respect to the wave propagation along the axis $z$. The number $\ell$ is referred to as the topological charge. As an eigenstate of the angular momentum operator $\hat{L}_{z}=-i\hbar\frac{\partial}{\partial\phi}$ , vortex waves carry a well-defined angular momentum of $\ell\hbar$ per photon or electron\cite{Allen1992}. Considering the transverse confinement, the simplest twisted state produced in TEM can be expressed in the form of Bessel-Gaussian\cite{Caron1999,Hricha2005} with
\begin{equation}
\psi(\rho,z)\propto e^{-{\rho^2}/{w_{0}^{2}}}J_{|\ell|}(k_{\perp}\rho)e^{ik_{z}z}.
\label{Eq::Bessel}
\end{equation}
Here 
$J_{|\ell|}$ is the Bessel function of the first kind, $k_{\perp}$ is the transverse wave number, $z$ is the propagation distance, $k_{z}$ is the axial wave number and $w_{0}$ is the beam waist\cite{Siegman1986}. Additionally, for this converging beam, a propagation-dependent phase shift needs to be introduced\cite{Bliokh2012,Siegman1986}
\begin{equation}
\Theta_{\rm Gouy}=(|\ell|+1)\arctan(z/z_{R}),
\end{equation}
which is dependent on the OAM of twisted electrons, and $z_{R}$ is the Rayleigh length.

\begin{figure}[htb]
\centering
\includegraphics [scale=0.3]{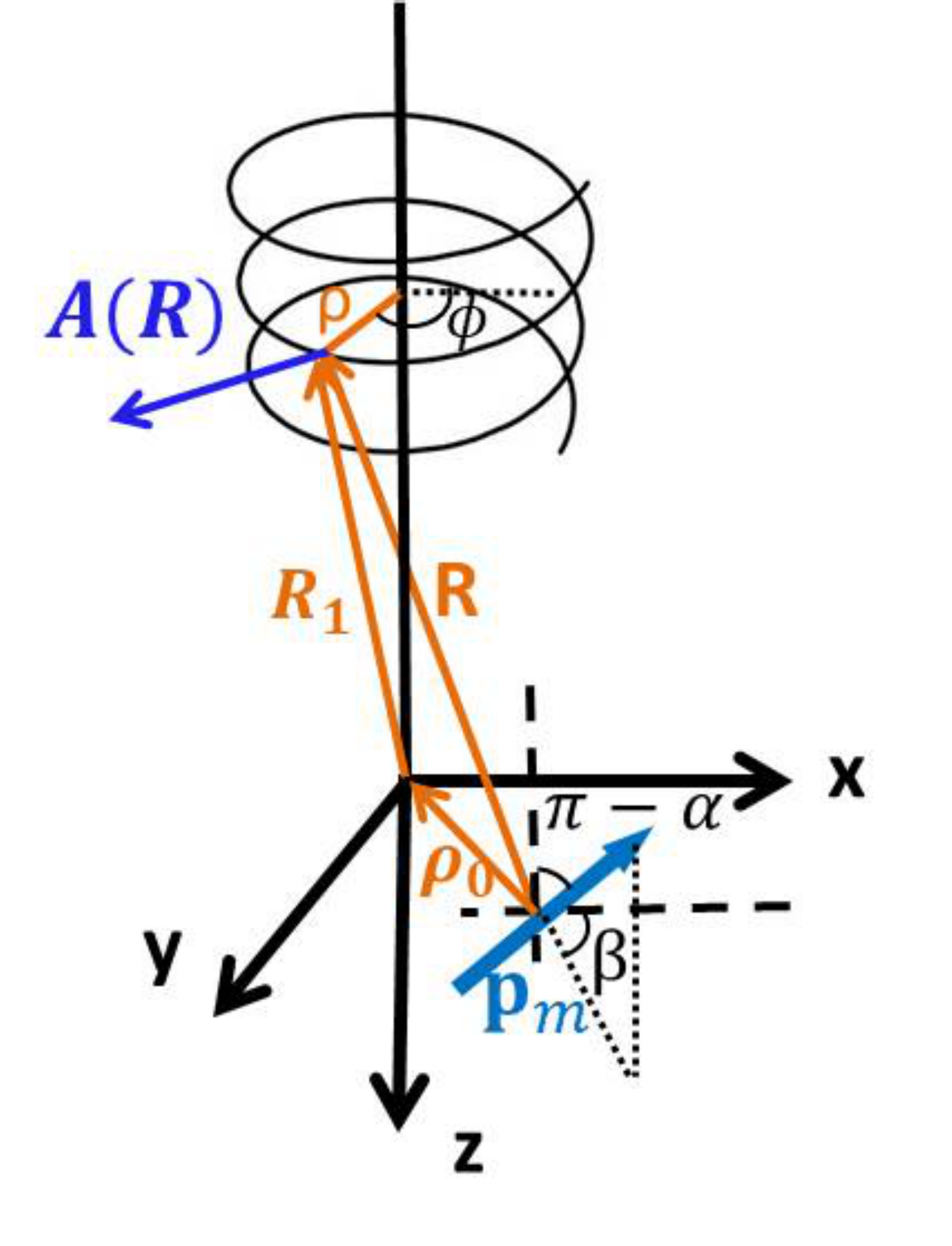}
\caption{The schematic diagram of a point magnetic dipole $\vec{p}_{m}$. $\alpha$ and $\beta$ describe the angles between the direction of the magnetic dipole moment and $z$ and $x$ axis, respectively. $\rho_{0}$ is the distance from the the propagation axis of the twisted electron beam. $\vec{A}(\vec{R})$ is the vector potential at position $\vec{R}$. }
\label{fig::vector potential}
\end{figure}

For the description of the fundamental, observable unit of magnetism, we start with the influence of magnetic dipole on the propagating twisted electrons. The influence of actual magnetic impurity on the electrons can then be calculated as the sum of individual effects of each magnetic dipole.  Let us consider a magnetic dipole located at the position $(x_{0},y_{0},0)$ with $\rho_{0}=\sqrt{x_{0}^{2}+y_{0}^{2}}$ representing the distance from the propagation axis of the twisted electron beam and $\alpha$ and $\beta$ describing the angles between the direction of the magnetic dipole moment and the $z$ and $x$ axis, respectively (see Fig.\ref{fig::vector potential}). Given a small transverse wave number $k_\perp$, we approximate the magnetic dipole by a point-like magnetic moment  $\vec{p}_{m}=p_{m}\left(\sin\alpha\cos\beta\hat{\vec{e}}_{x}+\sin\alpha\sin\beta\hat{\vec{e}}_{y}+\cos\alpha\hat{\vec{e}}_{z}\right)$, where $p_{m}$ is the magnitude of the dipole. The magnetic dipole model allows for calculating the magnetic strength of the magnet source. However, a restriction of the model is that the distance from the testing position to the magnet's position must be much larger than the size of the magnet. In our point magnetic moment case, the condition is fullfilled at any position in space. The vector potential associated with this dipole moment is given by the expression
\begin{equation}
\vec{A}(\vec{R})=\frac{\mu_{0}}{4\pi}\frac{\vec{p}_{m}\times\vec{R}}{R^{3}},
\label{Eq::vector potential_0}
\end{equation}
where $\vec{R}=\vec{\rho}_{0}+\vec{R}_{1}$ with $\vec{R}_{1}$ representing the position of twisted electrons,  $\mu_{0}=4\pi\times10^{-7}$ N/A$^{2}$ is the permeability of vacuum. The magnetic vector potential $\vec{A}$ in quantum physics has real, measurable effects. Here, it is directly related to the phase shifts under consideration.

Since the twisted electron beams are easily described in cylindrical coordinates, it is convenient to write the vector potential (Eq.\ref{Eq::vector potential_0}) in the same coordinate system:
\begin{eqnarray}
\vec{A}(\rho,\phi,z)&=&\frac{\mu_{0}}{4\pi}\frac{p_{m}}{\left[\left(\rho\cos\phi-x_{0}\right)^2+\left(\rho\sin\phi-y_{0}\right)^2+z^2\right]^{\frac{3}{2}}}
\nonumber \\
&&\Big\{\big[-z\sin\alpha\sin(\phi-\beta)+\cos\alpha(y_{0}\cos\phi-x_{0}\sin\phi)\big]\hat{e}_{\rho} \nonumber \\
&&+\big[\rho\cos\alpha-z\sin\alpha\cos(\phi-\beta)-\cos\alpha(y_{0}\sin\phi+x_{0}\cos\phi)\big]\hat{e}_{\phi} \nonumber \\
&&+\big[\rho\sin\alpha\sin(\phi-\beta)-\sin\alpha(y_{0}\cos\beta-x_{0}\sin\beta)\big]\hat{e}_{z}\Big\}
\label{Eq::vector potential_1}
\end{eqnarray}

Note that the vector potential has a spatial anisotropy. This means that the interaction of electrons and the magnetic dipole is dependent on the azimuthal angle $\phi$, which leads to the deformation of helical wavefront of the free twisted electron beams.

According to the phase object approximation (POA)\cite{Pozzi2014}, any electromagnetic field between planes $z_{i}$ and $z_{i+1}$ along the optical axis $z$ can be treated as a thin phase object. The corresponding transmission function $T$ can be formulated as
\begin{eqnarray}
T(\rho,\phi,z)&=&exp\Big[i\frac{\pi}{\lambda_{dB}}\frac{e}{E}\int_{z_{i}}^{z_{i+1}}V(\rho,\phi,z)dz-i\frac{e}{\hbar}\int_{z_{i}}^{z_{i+1}}A_{z}(\rho,\phi,z)dz\Big],
\label{Eq::POA}
\end{eqnarray}
where the incident electron beam direction $z$ is aligned with the optical axis, $V(\rho,\phi,z)$ is the electrostatic potential, $A_{z}(\rho,\phi,z)$ is the $z$ component of the magnetic vector potential and $e$, $\lambda_{dB}$, $\hbar$ and $E$ are the absolute values of the electron charge, the de Broglie electron wavelength, the reduced Planck constant and the kinetic energy of electrons in the non-relativistic approximation, respectively. The integral is extend along the $z$ axis, within the limits that include all of the field effects.

From Eqs. (\ref{Eq::vector potential_1}) and (\ref{Eq::POA}) it follows that, for a magnetic dipole, the direct integration along the $z$ axis between $-d$ and $d$ ($d>>\rho_{m}$) can be used to evaluate the electron-optical phase shift:
\begin{eqnarray}
\Phi_{z}(\rho,\phi)&=&-\frac{e}{\hbar}\int_{-d}^{d}A_{z}(\rho,\phi,z) dz  \nonumber \\
&=&-\frac{e}{\hbar}\frac{\mu_{0}}{4\pi}p_{m}\frac{2d}{\sqrt{\rho^{2}+d^2}}\sin\alpha
\frac{\rho\sin(\phi-\beta)-(y_{0}\cos\beta-x_{0}\sin\beta)}{\left(\rho\cos\phi-x_{0}\right)^2+\left(\rho\sin\phi-y_{0}\right)^2
}. \nonumber \\
\label{Eq::phi_z_1}
\end{eqnarray}

Here, $d$ is the distance between the electrons and the magnetic dipole along the $z$ axis, within which the vector potential has a significant effect on the phase $\Phi_{z}$. $A_{z}$ is the $z$-component of $\vec{A}$, which is related to the azimuthal angle of propagating electrons and the magnitude and orientation of the magnetic dipole. Hence, Eq.(\ref{Eq::phi_z_1}) describes an anisotropy phase difference that depends on the $\vec{p}_{m}$ and azimuthal angle $\phi$.

On the other hand, the Aharonov-Bohm phase is acquired by the twisted electrons when its path in a transverse plane encloses magnetic flux:
\begin{eqnarray}
\Phi_{\rm AB}&=& -\frac{e}{\hbar}\int_{0}^{2\pi} A_{\phi}\rho d\phi \nonumber \\
&=&-\frac{e}{\hbar}\frac{\mu_{0}}{4\pi}p_{m}\rho\cos\alpha\int_{0}^{2\pi}
\frac{\rho-(y_{0}\sin\phi+x_{0}\cos\phi)}
{\left[\left(\rho\cos\phi-x_{0}\right)^2+\left(\rho\sin\phi-y_{0}\right)^2\right]^{\frac{3}{2}}}d\phi.
\end{eqnarray}
$A_{\phi}$ is the $\phi$ component of the magnetic vector potential. This phase leads to either increase or decrease of the initial electron OAM state by $\nu=\Phi_{AB}/2\pi$, where $\nu$ is a real number.

In the transverse plane containing the dipole ($z=0$), after choosing the radius of a transverse loop as $\rho_{m}$, which corresponds to the radius of maximum intensity of beam, we obtain:
\begin{eqnarray}
\nu&=&-\frac{e}{\hbar}\frac{\mu_{0}}{8\pi^{2}}p_{m}\rho_{m}\cos\alpha\int_{0}^{2\pi}
\frac{\rho_{m}-(y_{0}\sin\phi+x_{0}\cos\phi)}
{\left[\left(\rho_{m}\cos\phi-x_{0}\right)^2+\left(\rho_{m}\sin\phi-y_{0}\right)^2\right]^{\frac{3}{2}}}d\phi.
\label{Eq::nu1}
\end{eqnarray}

The Gouy phase shift of the beam propagation from $-d/z_{R}<<-1$ to $d/z_{R}>>1$ is given by
\begin{equation}
\Theta_{\rm Gouy}=(|\ell+\nu|+1)\pi.
\label{Eq::Gouy Phase}
\end{equation}

Eq.(\ref{Eq::phi_z_1}) and Eq.(\ref{Eq::Gouy Phase}) show that the twisted electrons after passing through the magnetic dipole obtain both the usual optical phase and Gouy phase shift at the same time. Considering these two phases obtained by the twisted electron beam, the outgoing wave function of electrons at radius $\rho$ can be expressed by the following expression:
\begin{eqnarray}
\Psi(\vec{r})_{\ell}^{out}& \propto & e^{-\frac{\rho^2}{w_{0}^{2}}}J_{|\ell|}(k_{\perp}\rho)e^{i(\ell\phi+k_{z}z)}e^{i[\Theta_{\rm Gouy}+\Phi_{z}(\rho,\phi)]} \nonumber \\
&=& f(\rho,\phi)e^{ik_{z}z}.
\label{Eq::wave function}
\end{eqnarray}
\begin{figure*}[htb]
\centering
\includegraphics [scale=0.4]{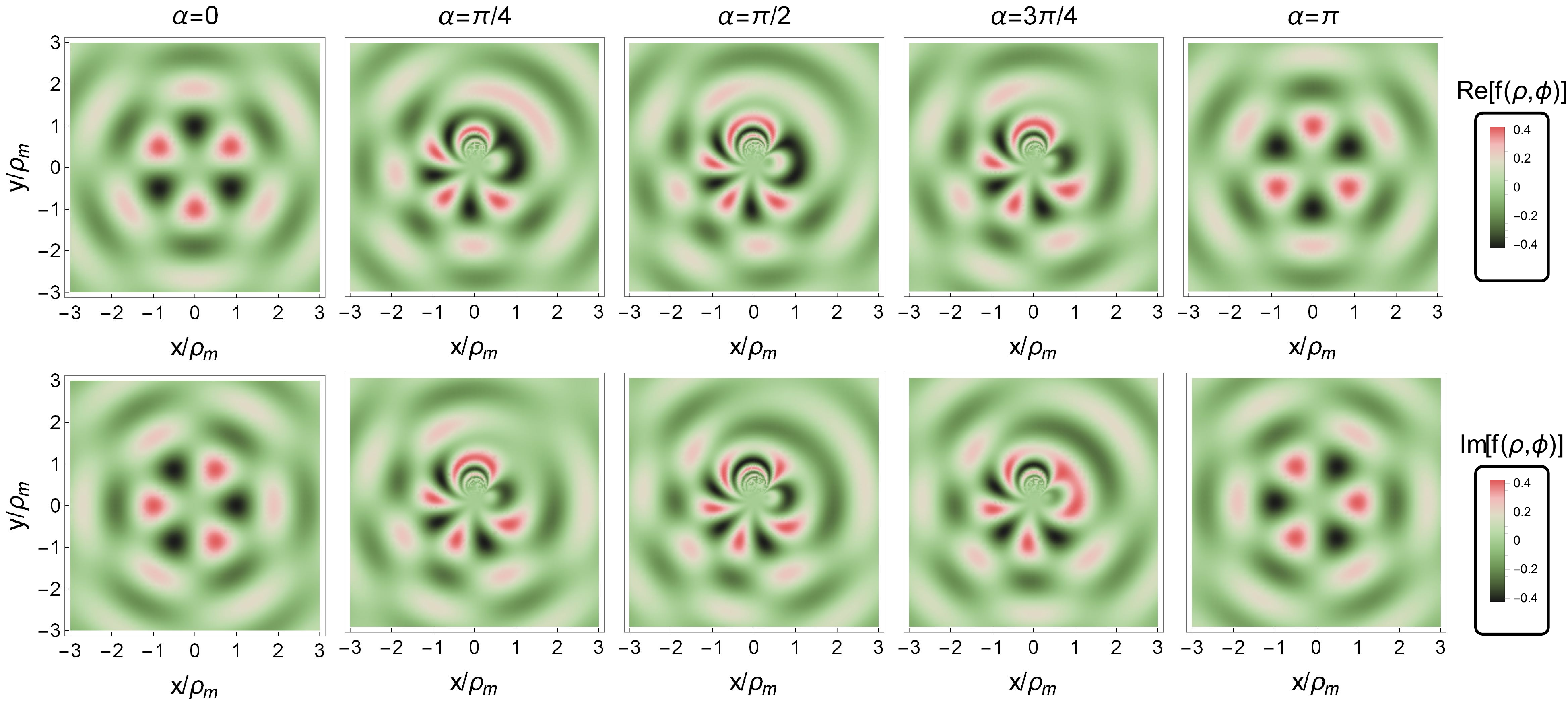}
\caption{The dependence of the real part (upper row) and the imaginary part (bottom row) of the outgoing wave function $f(\rho,\phi)$ on the angle $\alpha$ in the case $\beta=\pi/4$. The magnitude of magnetic dipole $p_{m}=2\times10^{-18}A\cdot m^{2}$. The kinetic energy of electrons is $300$ keV and  de Broglie wavelength $\lambda_{dB}=1.97pm$, $k_{\perp}=0.01k$, the OAM $\ell=3$, the waist of beam $w_{0}=400\lambda_{dB}$, $x_{0}=0.1\rho_{m}$ and $y_{0}=0.3\rho_{m}$.}
\label{fig::wavefunction1}
\end{figure*}
\begin{figure*}[htb]
\centering
\includegraphics [scale=0.4]{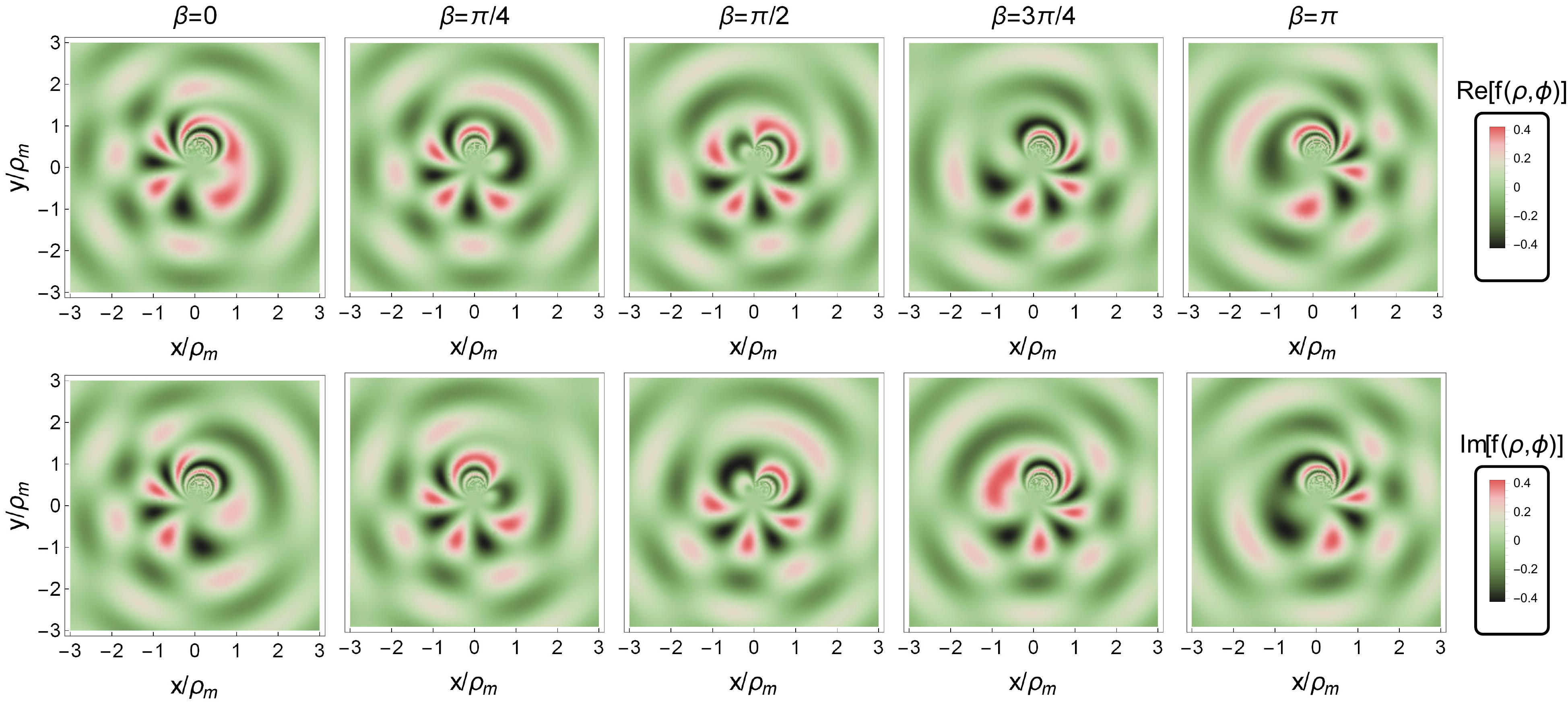}
\caption{The dependence of the real part (upper row) and the imaginary part (bottom row) of the outgoing wave function $f(\rho,\phi)$ on the angle $\beta$ in the case $\alpha=\pi/4$. Other parameters are the same as specified in the caption of Fig.\ref{fig::wavefunction1}.}
\label{fig::wavefunction2}
\end{figure*}

The wave function $f(\rho,\phi)$ in the plane perpendicular to the propagation axis for different $\alpha$ and $\beta$ angles is shown in Fig.\ref{fig::wavefunction1} and Fig.\ref{fig::wavefunction2}, respectively. The period of the wave function is $\frac{2\pi}{\ell}$ when $\alpha=0,\pi$, however, the $\phi$-dependent optical phase in other cases ($\alpha\ne 0,\pi$) affects the wave function.

It's important to note that when the magnetic dipole is located very near the propagation axis of electrons, the vector potential can be expressed in more concise form:
\begin{eqnarray}
\vec{A}(\rho,\phi,z)&=&\frac{\mu_{0}}{4\pi}\frac{p_{m}}{(\rho^{2}+z^{2})^{3/2}}\Big\{-z\sin\alpha\sin(\phi-\beta)\hat{\vec{e}}_{\rho}
\nonumber \\
&&+\left[\rho\cos\alpha-z\sin\alpha\cos(\phi-\beta)\right]\hat{\vec{e}}_{\phi}\nonumber \\
&&+\rho\sin\alpha\sin(\phi-\beta)\hat{\vec{e}}_{z} \Big\}.
\label{Eq::vector potential_2},
\end{eqnarray}
so that
\begin{eqnarray}
\Phi_{z}(\rho,\phi)&=&-\frac{e}{\hbar}\frac{\mu_{0}}{4\pi}p_{m}\frac{2d}{\rho\sqrt{\rho^{2}+d^2}}\sin\alpha\sin(\phi-\beta).
\label{Eq::phi_z_2}
\end{eqnarray}
The change of the electron's OAM $\nu$, according to Eq.\ref{Eq::nu1}, has the form:
\begin{eqnarray}
\nu&=&-\frac{e}{\hbar}\frac{\mu_{0}}{4\pi}p_{m}\frac{\cos\alpha}{\rho_{m}}.
\end{eqnarray}

\section{Phase shifts visualization}

\begin{figure*}[htb]
\centering
\includegraphics [scale=0.4]{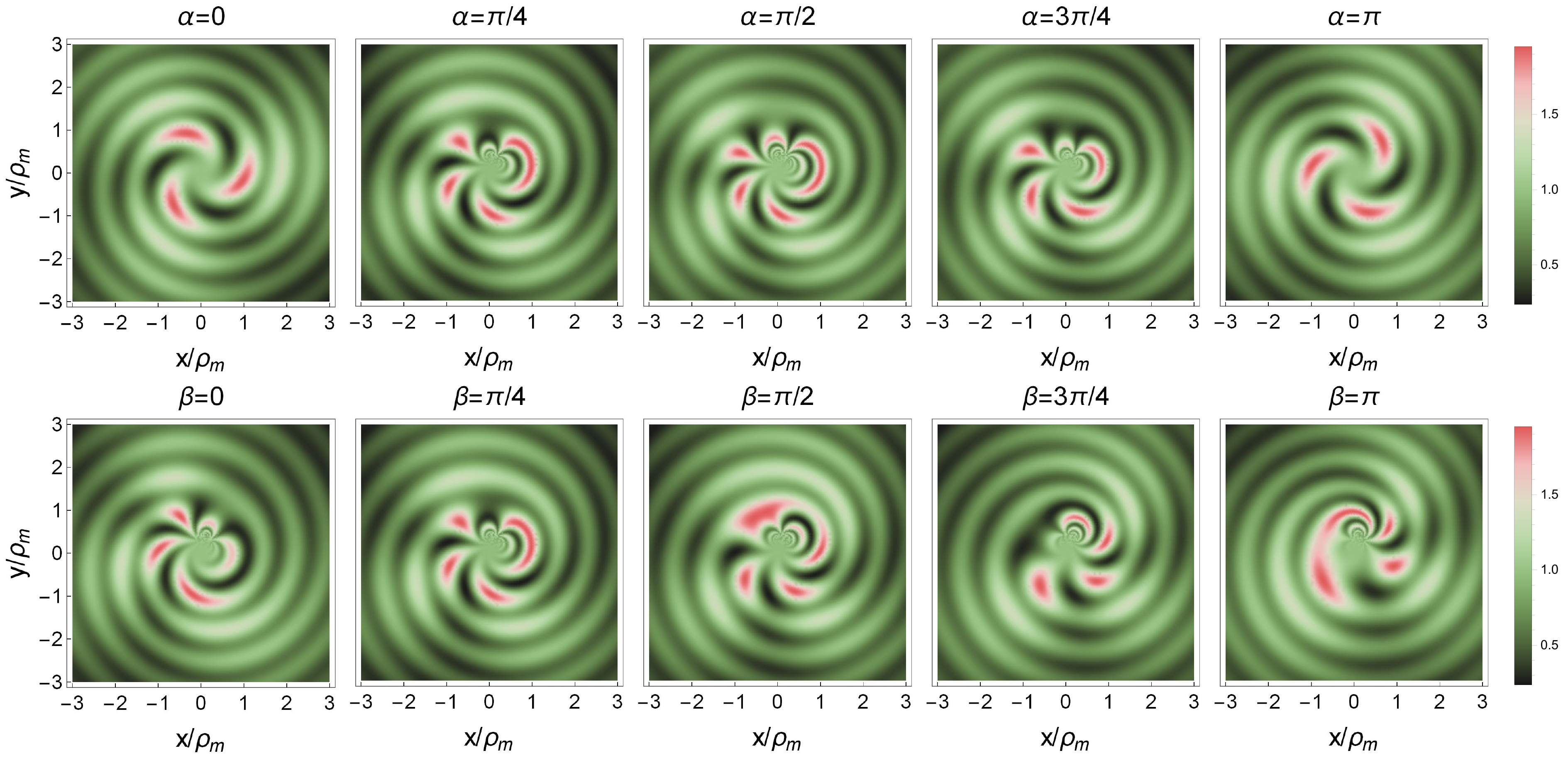}
\caption{Interference patterns of the outgoing twisted electron beams and a reference Gaussian-cylindrical wave for (upper row) different $\alpha$ ($\beta=\pi/4$) and (bottom row) $\beta$ ($\alpha=\pi/4$) angles, respectively. The wave number $k^{\prime}=k_{\perp}$ and other parameters are the same as given in Fig.\ref{fig::wavefunction1}.}
\label{fig::interference1}
\end{figure*}

\begin{figure*}[htb]
\centering
\includegraphics [scale=0.4]{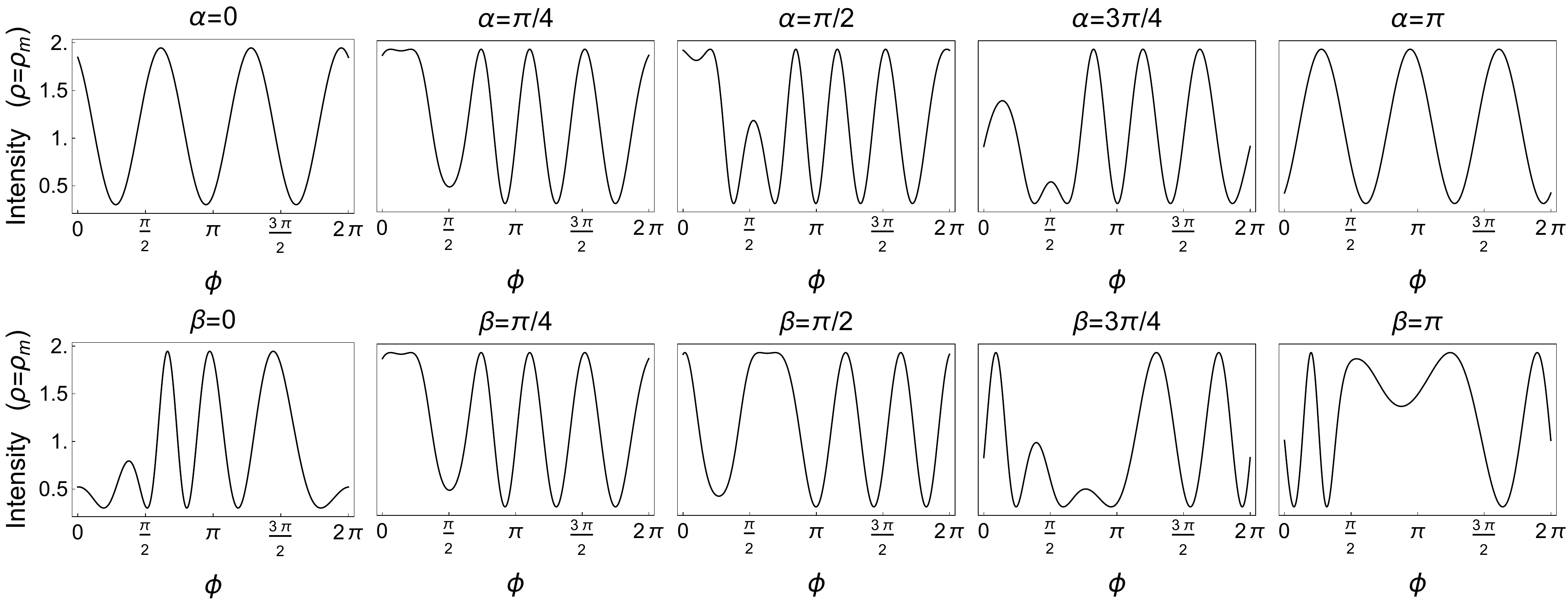}
\caption{The corresponding distributions of the interference intensity (see Fig.\ref{fig::interference1}) with respect to the azimuthal angle $\phi$ at the radius of maximum intensity of beam.}
\label{fig::intensity1}
\end{figure*}

\begin{figure*}[htb]
\centering
\includegraphics [scale=0.4]{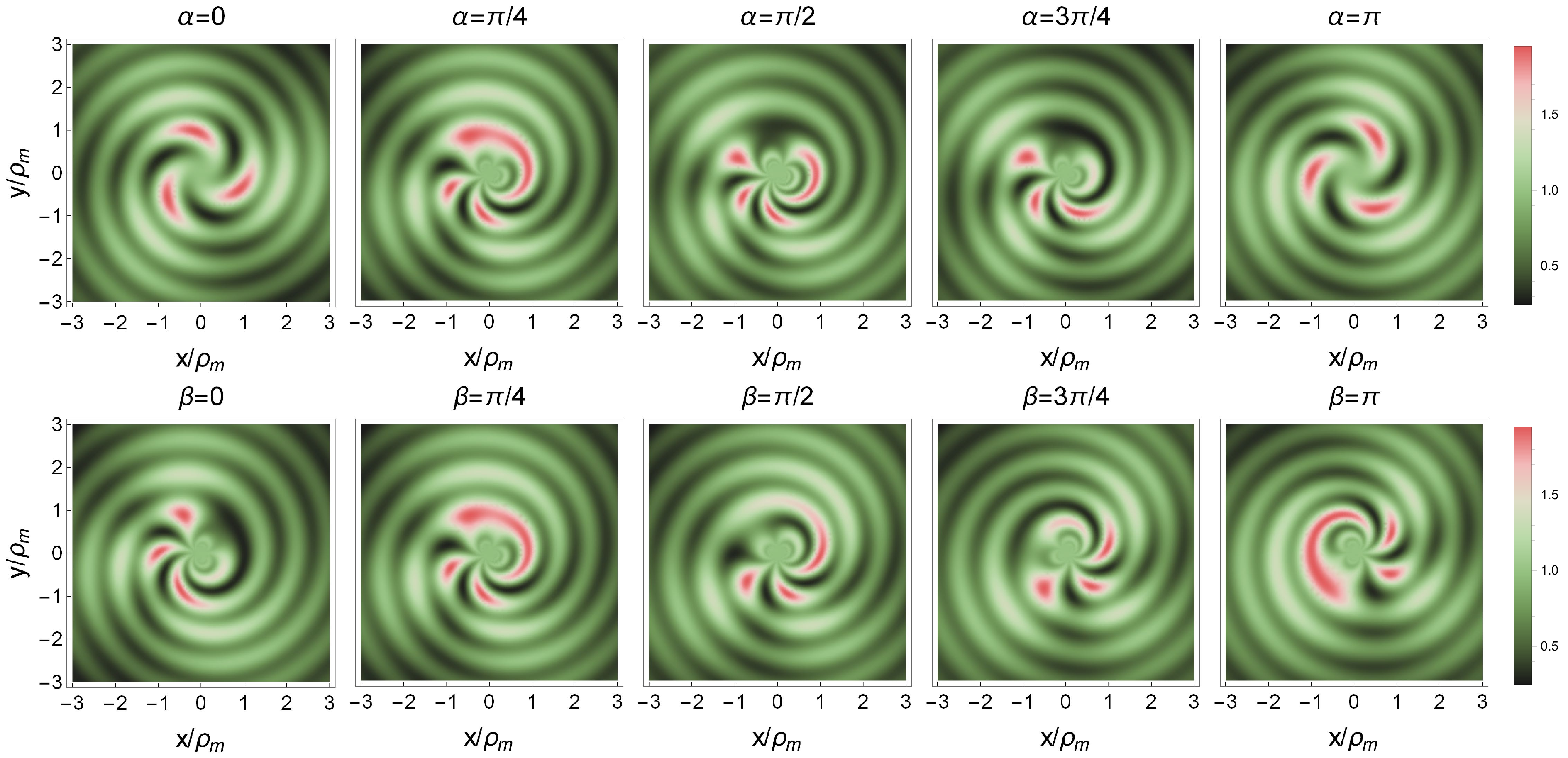}
\caption{Interference patterns corresponding to a particular case of Fig.\ref{fig::interference1} that $\rho_{0}=0$. }
\label{fig::interference2}
\end{figure*}

In order to visualize the effect of the magnetic dipole on the twisted electrons, we interfered the outgoing twisted electron beam with a reference Gaussian cylindrical wave $\psi_{ref}\propto e^{-{\rho^2}/{w_{0}^{2}}}e^{ik^{\prime}\rho}$. $k^{\prime}$ is the wave number in the transverse plane of the reference wave.  Fig.\ref{fig::interference1} and Fig.\ref{fig::interference2} show how the interference pattern of the outgoing twisted electron wave and cylindrical wave varies with the $\alpha$ and $\beta$ angles when the magnetic dipole is off and on the propagation axis, respectively. The corresponding distributions of the interference intensity with respect to the azimuthal angle $\phi$ at $\rho=\rho_{m}$ ( from Fig.\ref{fig::interference1}) is demonstrated in Fig.\ref{fig::intensity1}. The magnitude of the dipole moment $p_{m}=2\times 10^{-18} A\cdot m^{2}$. These interference patterns represent that the phases obtained by the twisted electrons during propagation through the magnetic dipole. When the direction of the magnetic dipole moment is along the $z$ axis ($\alpha=0,\pi$), the electron-optical phase shift $\Phi_{z}$ becomes zero and only the Gouy phase exists. In this case, the helical wavefront is not deformed and the patterns are independent of the angle $\beta$ but rotated by an angle ${\Theta_{\rm Gouy}}/{\ell}$, which is related to the magnitude of the magnetic dipole. However, in more general case ($\alpha\neq 0,\pi$),
the two phase shifts mentioned above exist at the same time and the patterns not only rotate but also deform when the angle $\alpha$ or $\beta$ changes, as shown in  Fig.\ref{fig::interference1} and Fig.\ref{fig::interference2}. For a better illustration, we calculated the $\phi$-resolved interference intensity as a function of the azimuthal angle $\phi$ at radius $\rho=\rho_{m}$, and the obtained results are shown in Fig.\ref{fig::intensity1}. The width of the interference intensity peak at half-height can be roughly considered as the size of the bright spots, and the center position of the spots can be obtained directly as well. Clearly, the interference pattern rotates and deforms when the direction of the dipole moment changes.


\begin{figure}[htb]
\centering
\includegraphics [scale=0.45]{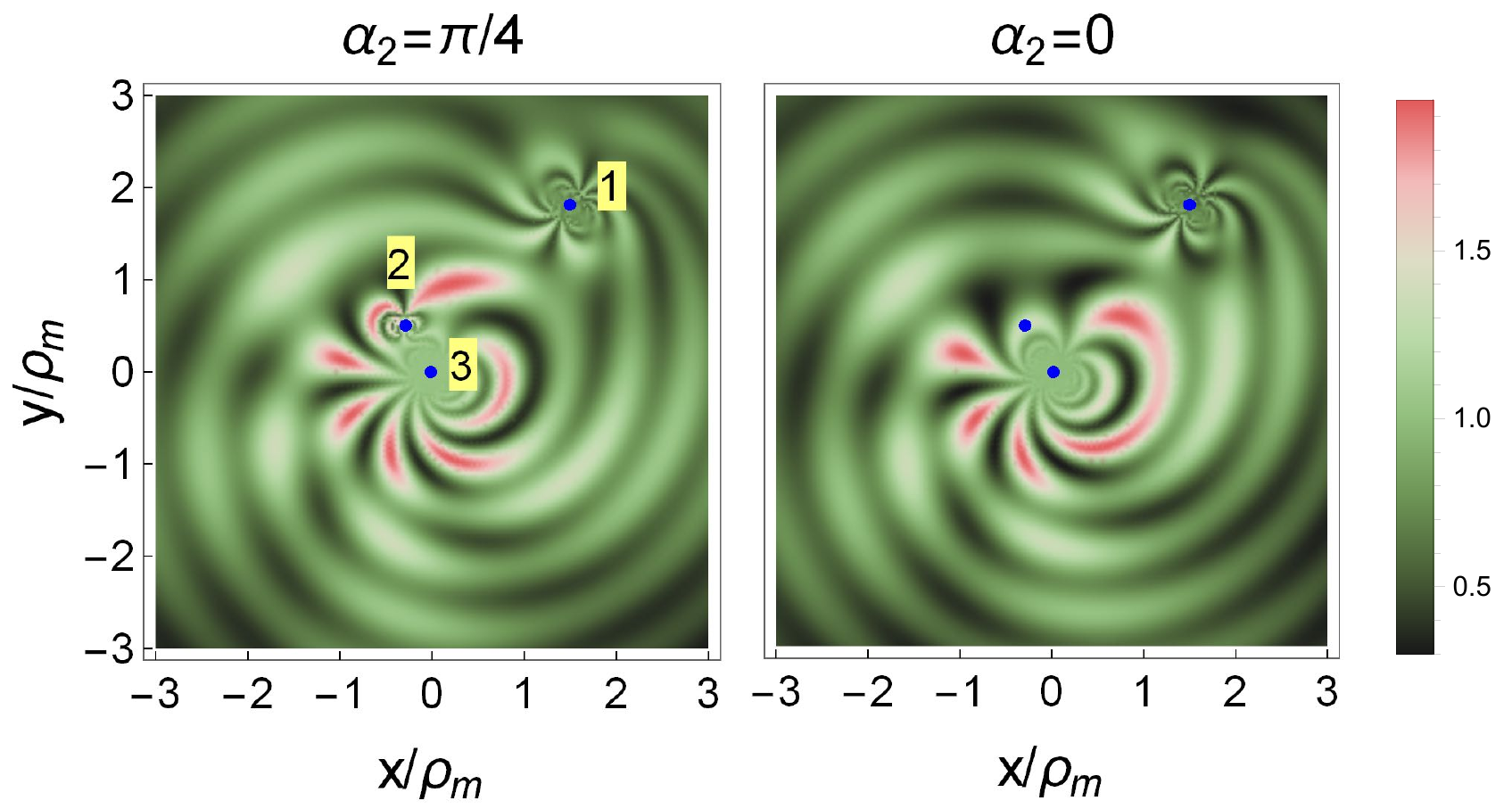}
\caption{Interference patterns of the outgoing twisted electron beams and a reference wave under the influence of three magnetic dipoles. The strength of magnetic dipoles is $2\times 10^{-18}$, $1\times 10^{-18}$ and $3\times 10^{-18}A\cdot m^{2}$. $\alpha_{1}=\pi/3$, $\beta_{1}=\pi/4$; $\alpha_{2}=\pi/4,0$, $\beta_{2}=\pi/2$; $\alpha_{3}=2\pi/3$, $\beta_{3}=\pi/4$. $a_{1}=1.5$, $b_{1}=1.8$, $a_{2}=-0.3$, $b_{2}=0.5$, $a_{3}=b_{3}=0$. The other parameters are the same as in Fig.\ref{fig::wavefunction1}. }
\label{fig::interference3}
\end{figure}

The magnetic feature of the magnetic impurity can be considered as an array of structural localized magnetic dipoles possessing different strengths. Taking three dipoles as an example, these dipoles locate at $(a_{1}\rho_{m},b_{1}\rho_{m},0)$, $(a_{2}\rho_{m},b_{2}\rho_{m},0)$ and $(a_{3}\rho_{m},b_{3}\rho_{m},0)$ respectively and the corresponding strength and orientation are $p_{m1}(\sin\alpha_{1}\cos\beta_{1}\hat{\vec{e}}_{x}+\sin\alpha_{1}\sin\beta_{1}\hat{\vec{e}}_{y}+\cos\alpha_{1}\hat{\vec{e}}_{z})$, $p_{m2}(\sin\alpha_{2}\cos\beta_{2}\hat{\vec{e}}_{x}+\sin\alpha_{2}\sin\beta_{2}\hat{\vec{e}}_{y}+\cos\alpha_{2}\hat{\vec{e}}_{z})$ and $p_{m3}(\sin\alpha_{3}\cos\beta_{3}\hat{\vec{e}}_{x}+\sin\alpha_{3}\sin\beta_{3}\hat{\vec{e}}_{y}+\cos\alpha_{3}\hat{\vec{e}}_{z})$. Without loss of generality, let's take the angles $\alpha_{1}=\pi/3$, $\beta_{1}=\pi/4$; $\alpha_{2}=\pi/4,0$, $\beta_{2}=\pi/2$; $\alpha_{3}=2\pi/3$, $\beta_{3}=\pi/4$. The strength of magnetic dipoles is $2\times 10^{-18}$, $1\times 10^{-18}$ and $3\times 10^{-18}A\cdot m^{2}$. The interference pattern of the outgoing twisted electron wave and the reference wave are shown in Fig.\ref{fig::interference3}. The distribution of dipoles can be clearly seen in the figure, and they are marked by blue dots. Furthermore, the strength and orientation of each magnetic dipole can be obtained from the distribution of interference intensity with respect to the azimuthal angle at a certain radius $\rho$. For those dipoles whose orientation is parallel or anti-parallel to the propagation axis of the twisted electron beam (see pattern on the right of Fig.\ref{fig::interference3}), we can obtain their distribution by rotating the material at a small angle.

\section{Conclusion}
We investigated theoretically the influence of a magnetic impurity on the twisted electron beams generated by TEM. Considering the diversity of magnetic impurities, a more general magnetic dipole case was investigated. The vector potential of the magnetic dipole contributes to the electron optical phase and Gouy phase shifts. The Aharonov-Bohm phase of electrons causes the increase or decrease of the initial OAM, resulting in the Gouy phase shift. In addition, this phase is independent on the azimuthal angle $\phi$, so it only can rotate the interference pattern. Unlike the homogeneous Gouy phase, the electron optical phase shift arising from the $z$ component of the vector potential of the magnetic dipole is dependent on the azimuthal angle $\phi$ and, thus, becomes anisotropic, which leads to the deformed and rotational interference patterns. The $\phi$-resolved interference intensity is very sensitive to both the magnitude and the orientation of the magnetic dipole, of which the fitting function can be used to reveal the peculiarities of the dipole moment $\vec{p}_{m}$.  For the representative atomic- and nanoscale magnetic impurity, it was demonstrated that the impurity can be considered as a sum of magnetic dipoles and the detailed analysis of the superposition of interference patterns with different magnitude and direction of dipole moments would help in understanding the magnetic structure of the impurity.

\section*{Acknowledgements}
This work was supported by National Natural Science Foundation of
China (Grant No.$11847201$, No.$11975320$, and No.$11834005$), the National Key Research and Development Program of China $($No.$2016$YFE$0130800)$ and the Open Project of Key Laboratory for Magnetism and Magnetic Materials of the Ministry of Education, Lanzhou University (LZUMMM$2018016$ and LZUMMM$2019011$).


\section*{References}
\begin{thebibliography}{}
\bibitem{Bliokh_2007} K. Y. Bliokh, Y. P. Bliokh, S. Savel'ev, and F. Nori 2007 \textit{Phys. Rev. Lett.} \textbf{99} 190404.
\bibitem{Uchida_2010} M. Uchida and A. Tonomura 2010 \textit{Nature} \textbf{464} 737-739.
\bibitem{Verbeeck_2010} J. Verbeeck, H. Tian, and P. Schattschneider 2010 \textit{Nature} \textbf{467} 301-304.
\bibitem{Mcmorran_2011} B. J. Mcmorran and J. Unguris 2011 \textit{Science} \textbf{331} 192-195.
\bibitem{Schattschneider_2011} P. Schattschneider and J. Verbeeck 2011 \textit{Ultramicroscopy} \textbf{111} 1461-68.
\bibitem{Bliokh_2011} K. Y. Bliokh, M. R. Dennis, and F. Nori 2011 \textit{Phys. Rev. Lett.} \textbf{107} 174802.
\bibitem{Ivanov_2011} I. P. Ivanov 2011 \textit{Phys. Rev. D} \textbf{83} 093001.
\bibitem{Karimi_2012} E. Karimi, L. Marrucci, V. Grillo, and E. Santamato 2012 \textit{Phys. Rev. Lett.} \textbf{108} 044801.
\bibitem{Lloyd_2012} S. Lloyd, M. Babiker, and J. Yuan 2012 \textit{Phys. Rev. Lett.} \textbf{108} 074802.
\bibitem{Gallatin_2012} G. M. Gallatin and B. McMorran 2012 \textit{Phys. Rev. A} \textbf{86} 012701.
\bibitem{RHarvey2014}  T. R Harvey, J. S Pierce, A. K Agrawal, P. Ercius, M. Linck, and B.J. McMorran 2014 \textit{New J. Phys.} \textbf{16} 093039.
\bibitem{Ju2018} L.B. Ju, C.T. Zhou, K. Jiang, T.W. Huang, H. Zhang, T.X. Cai, J.M. Cao, B. Qiao, and S.C. Ruan 2018 \textit{New J. Phys.} \textbf{20} 063004.
\bibitem{Pierce2019} J. Pierce, J. Webster, H. Larocque, E. Karimi, B. McMorran, and A. Forbes 2019 \textit{New J. Phys.} \textbf{21} 043018.

\bibitem{Mafakheri2017} E. Mafakheri, A.H. Tavabi, P.H. Lu, R. Balboni, F. Venturi, C. Menozzi, G.C. Gazzadi, S. Frabboni, A. Sit, R.E. Dunin-Borkowski, E. Karimi, and V. Grillo 2017 \textit{Appl. Phys. Lett.} \textbf{110} 093113.
\bibitem{Grillo2017a} V. Grillo, T.R. Harvey, F. Venturi, J.S. Pierce, R. Balboni, F. Bouchard, G.C. Gazzadi, S. Frabboni, A.H. Tavabi, Z.A. Li, R.E. Dunin-Borkowski, R.W. Boyd, B.J. McMorran, and E. Karimi 2017 \textit{Nat. Commun.} \textbf{8} 689.
\bibitem{Bliokh2017a} K.Y. Bliokh, I.P. Ivanov, G. Guzzinati, L. Clark, R. Van Boxem, A. B\'{e}ch\'{e}, R. Juchtmans, M.A. Alonso, P. Schattschneider, F. Nori, and J. Verbeeck 2017 \textit{Phys. Rep.} \textbf{690} 1-70.
\bibitem{Lloyd2017a} S.M. Lloyd, M. Babiker, G. Thirunavukkarasu, and J. Yuan 2017 \textit{Rev. Mod. Phys.} \textbf{89} 035004.
\bibitem{Gnanavel2012} T. Gnanavel, J. Yuan, and M. Barbiker 2012 \textit{Proceedings of the European Microscopy Congress}.
\bibitem{Bliokh2012} Bliokh, K. Y., Schattschneider, P., Verbeeck, J., and Nori, F. 2012 \textit{Phys. Rev. X} \textbf{2} 041011.
\bibitem{Ivanov2013} I.P. Ivanov and D. V. Karlovets 2013 \textit{Phys. Rev. Lett.} \textbf{110} 264801.
\bibitem{Kosheleva2018} V.P. Kosheleva, V.A. Zaytsev, A. Surzhykov, V.M. Shabaev, and T. St\"{o}hlker 2018 \textit{Phys. Rev. A} \textbf{98} 022706.
\bibitem{Maiorova2018}  A. V. Maiorova, S. Fritzsche, R.A. M\"uller, and A. Surzhykov 2018 \textit{Phys. Rev. A} \textbf{98} 042701.
\bibitem{Guzzinati2013}  G. Guzzinati, P. Schattschneider, K.Y. Bliokh, F. Nori, and J. Verbeeck 2013 \textit{Phys. Rev. Lett.} \textbf{110} 093601.
\bibitem{Kondo effect} A. C. Hewson 1997 \textit{The Kondo Problem to Heavy Fermions (Cambridge University Press)}.
\bibitem{Aharonov1959} Aharonov, Y, and D Bohm. 1959 \textit{Phys. Rev} \textbf{115} 485.
\bibitem{Siegman1986} Siegman, A. E. 1986 \textit{Lasers(University Science Books)} \textbf{37},462-466.
\bibitem{Nye1974} J.F. Nye and M.V. Berry 1974 \textit{Proc. R. Soc. London A} \textbf{336} 165-190.
\bibitem{Allen1992} L. Allen, M.W. Beijersbergen, R.J.C. Spreeuw, and J.P. Woerdman 1992 \textit{Phys. Rev. A} \textbf{45} 8185-89.
\bibitem{Caron1999} C.F.R. Caron and R.M. Potvliege 1999 \textit{Opt. Commun.} \textbf{164}, 83-93.
\bibitem{Hricha2005} Z. Hricha and A. Belafhal 2005 \textit{Opt. Commun.} \textbf{255} 235-240.
\bibitem{Pozzi2014} G. Pozzi, M. Beleggia, T. Kasama, and R.E. Dunin-borkowski 2014 \textit{Comptes Rendus Phys.} \textbf{15} 126-139.
\end{thebibliography}
\end{document}